\documentclass[12pt,a4pape]{article}
\usepackage{amsfonts}
\usepackage{amsmath,float}
\usepackage{graphicx}
\usepackage{color}
\usepackage{physics}

		\newcommand{\der}[3]{\frac{d^#3 #1}{d#2^#3}}
		\newcommand{\deruno}[2]{\frac{d#1}{d#2}}
		\newcommand{\pder}[3]{\frac{\partial ^#3 #1}{\partial #2^#3}}

\newcommand{\Ref}[1]{(\ref{#1})}
\begin{document}
\title{Vacuum energy for generalised Dirac combs at $T=0$}
\author{M. Bordag$^{1}$\thanks{bordag@uni-leipzig.de}, J. M. Mu$\tilde{\rm n}$oz-Casta$\tilde{\rm n}$eda$^{2}$\thanks{jose.munoz.castaneda@uva.es} and\\
 L. Santamar\'\i  a-Sanz$^{2}$\thanks{lucia.santamaria@uva.es}\\
\small $^{1}$Institut f{\"u}r Theoretische Physik, Universit{\"a}t Leipzig (GERMANY)\\
\small $^{2}$Departamento de F{\'\i}sica Te{\'o}rica At{\'o}mica y {\'O}ptica, Universidad de Valladolid (SPAIN)}

\date{\today}

\maketitle

\begin{abstract}
The quantum vacuum energy for a hybrid comb of Dirac $\delta$-$\delta^\prime$ potentials is computed by using the energy of the single $\delta$-$\delta^\prime$ potential over the real line that makes up the comb. The zeta function of a comb periodic potential is the continuous sum of zeta functions over the dual primitive cell of Bloch quasi-momenta. The result obtained for the quantum vacuum energy is non-perturbative in the sense that the energy function is not analytical for small couplings 
\end{abstract}\thispagestyle{empty}

\section{Introduction}
In this paper we analyse a generalisation of the Kronig-Penney model \cite{kronig-prsa31} in which the periodic point potential considered is a combination of the Dirac $\delta$-potential and its first derivative, i.e.  the $\delta$-$\delta'$ potential (see ref. \cite{GNN}).
Note that the Kronig-Penney model is an example of a one dimensional exactly solvable periodic potential, widely used in Solid State Physics to describe electrons moving in an infinite periodic array of rectangular potential barriers. The $\delta$-$\delta'$ potential has been a focus of attention over the last few years \cite{MM,gadella-jpa16,FAS,gad-ijtp11,gad-ijtp11-2}, but the $\delta$-$\delta'$ comb as a classical background in interaction with a scalar quantum field has not been considered.

%jpa-ggmn

The main goal of this work is to compute the quantum vacuum energy of a scalar field propagating in a (1+1)-dimensional spacetime  in interaction with the background of a generalised Dirac comb composed of $\delta$-$\delta'$ potentials, see \cite{GNN,MM,gadella-jpa16}. Interpreting the scalar field as electrons (disregarding spin) we would get a (non additive) contribution to the internal energy of the lattice. In a periodic structure it is possible to calculate the quantum vacuum energy per unit cell, which gives a contribution to the internal pressure of the lattice. In addition, it is possible to interpret the quantum scalar field as phonons of the lattice. In such a case we would obtain the phonon contribution to the internal pressure of the lattice when computing the quantum vacuum energy per unit cell. However, since the (1+1)-dimensional quantum field theory is a highly simplified theoretical model  we will not go into more detail about the interpretation.

Specifically, we study the one dimensional periodic Hamiltonian
\begin{equation}\label{1}
\mathcal{H}=-\frac{\hbar^2}{2m}\der{}{x}{2} + V(x) \qquad \textrm{where} \qquad V(x)=\sum_{n\in \mathbb{Z}} \mu \delta(x-nd)+ 2\lambda \delta^{\prime}(x-nd),
\end{equation}
with couplings $\mu, \lambda \in \mathbb{R}$, and lattice spacing $d>0$.
We will work with dimensionless quantities defined as
\begin{eqnarray}
y=\frac{mc}{\hbar} x, \qquad a=\frac{mc}{\hbar}d, \qquad w_0= \frac{1}{\hbar c} \mu, \qquad w_1= \frac{m}{\hbar^2} \lambda,
\end{eqnarray}
so that $[y,a,w_0,w_1]=1$. In that way, the dimensionless time independent Schr\"odinger  equation for the one-particle states of a quantum scalar field is
\begin{equation}
\left(-\pder{}{y}{2}+V(y)\right)\phi(y)=k^2 \phi(y),\qquad V(y)=\sum_{n\in \mathbb{Z}} w_0 \delta(y-na)+ 2w_1 \delta^{\prime}(y-na).\label{hamilt}
\end{equation}
Its solutions enables us to determine the energy levels and energy bands of the crystal. Following Ref. \cite{course-comb} the general form of the band equation in terms of scattering coefficients ($t,r_R,r_L$) for the compact supported potential from which the comb is built is
\begin{equation}
cos (qa)=\frac{e^{iak}(t(k)^2-r_R(k)r_L(k))+e^{-iak}}{2t(k)}\label{band-eq},
\end{equation}
being $q$ the quasi-momentum.
This equation relates the quasi-momenta $q \in[-\pi/a,\pi/a]$ in the first Brillouin zone and the wave-vector $k$. The quasi-momentum determines the Bloch periodicity for a given wave function on the lattice:
\begin{equation}
\phi(y+a)=e^{iqa}\phi(y).
\end{equation}
Since the cosine of the left hand side of \eqref{band-eq} is a bounded function, the energy spectrum of the system is organized into allowed/forbidden energy bands/gaps. As a particular case, when the scattering data for a Dirac-$\delta$ potential $V=w_0\delta(x)$ on the line \cite{JMG}
\begin{equation}
t_\delta(k)=\frac{2 i k}{2ik-w_0},\quad r_\delta(k)=\frac{w_0}{2ik-w_0}
\end{equation} 
are plugged into equation \eqref{band-eq} we obtain 
\begin{equation}
\cos(qa)=\cos(k a)+\frac{w_0}{2 k}\sin(k a)
\end{equation}
which is the well known band equation for the Kronig-Penney model \cite{kronig-prsa31}.

The general secular equation \eqref{band-eq} will enable us to calculate the vacuum energy of the crystal.
The vacuum energy per unit cell (in the interval $[0,a]$) is computed by spatially integrating the expectation value of the $ 00$-component of the energy-momentum tensor $T_{\mu\nu}$:
\begin{equation}
E_0= \int_0^a dy \bra{0}T_{00}\ket{0}.
\end{equation}
The non regularised infinite quantum vacuum energy can be represented as well as the summation over modes of the spectrum corresponding to the one-particle states of the field theory. 
\begin{equation}
E_0= \int_0^a dy \bra{0}T_{00}\ket{0}=\sum_{n}k_n
\end{equation}
being $\{\omega_n^2=k_n\}$ the eigenvalues characterising the one-particle states of the quantum field theory given by the equation \eqref{band-eq}. 
The ultraviolet divergences that appear naturally in this expression must be subtracted taking into account the self-energy of the individual potential that makes up the comb and the fluctuations of the field in the chosen background. The calculation of $\bra{0}T_{00}\ket{0}$ provides the energy density per unit length within a unit cell. This of course contains much more information than just the total energy contained in a unit cell. Nevertheless, the calculation using Green functions will not be addressed in this paper for most general combs. On the other hand we can compute $E_0$ using spectral zeta functions \cite{spectral} to skip the intermediate calculation of $\bra{0}T_{00}\ket{0}$ for which the exact Green function of the quantum field on the crystal is needed. When using the zeta function approach the infinite contributions are subtracted using the regularised expression for the quantum vacuum energy:
\begin{equation}
E_0(s)=\sum_{n}k_n^{-s}.\label{e-zeta}
\end{equation}
This expression is nothing but the spectral zeta function associated to the Sch{\"o}dinger operator defined in Eq. \eqref{hamilt}. In order to subtract the divergences one has to perform the analytic continuation of \eqref{e-zeta} for $s$ to the whole complex plane, and then subtract the contribution of the pole at $s=-1$. A detailed explanation of how to proceed in most general cases is explained in Refs. \cite {spectral,vassilevich-physrept03,emil-book}.

The structure of the present paper is the following. In section 2 we reproduce some basic results on spectral zeta functions that are needed throughout the paper. The section 3 provides a way to re-interpret a general comb formed by superposition of identical potentials with compact support centered at the lattice points, as a 1-parameter family of pistons mimicked by quasi-periodic boundary conditions using the formalism to characterise selfadjoint extensions developed in \cite{aso-mc2}. Afterwards in section 4 and subsection 5.1, we will use the results from Refs. \cite{aso-mc2,mc-kk1} to give a general formula for the finite quantum vacuum energy general comb formed by superposition of identical potentials with compact support. The subtraction of infinites follows from \cite{aso-mc2}. The rest of section 5 is dedicated to the numerical results for the particular example of the $\delta$-$\delta'$ comb, and the non-perturbative character inherent to the quantum vacuum energy of this particular example. Finally in section 6 we explain the conclusions of our paper.

\section{Some basics formulas on spectral zeta functions}
In general, given an arbitrary potential with {\it small}\footnote{Many of the results of this paper generalise straightforward to any comb built from a superposition of potentials with compact support centered at the lattice points, provided that the compact support of such potentials is smaller than the lattice spacing.} support and its associated comb, the secular equation \eqref{band-eq} can not be solved. Nevertheless the summation over eigenvalues in \eqref{e-zeta} can be rewritten down using the residue theorem. In this section we explain the method to replace the summation over eigenvalues in eq. \eqref{e-zeta} by a complex contour integral involving the logarithmic derivative of the function that defines the secular equation \eqref{band-eq}.

Let $\hat{H}$ be an elliptic non-negative selfadjoint, second order differential operator and $f_{\hat{H}}(z)$ an holomorphic function on $\mathbb{C}$ such that
 \begin{eqnarray*}
&&i) \lim_{z\rightarrow 0}  f_{\hat{H}}(z)\neq 0,\infty.	\nonumber\\
&& ii) \textrm{If we define} \nonumber\\
&& \qquad   \qquad  \qquad Z(f_{\hat{H}})\equiv \lbrace k_n \in \mathbb{R} / f_{\hat{H}}(k_n)=0\rbrace\nonumber\\
&&\qquad   \qquad  \qquad \tilde{\sigma}(\hat{H})\equiv \lbrace \lambda_n \in \mathbb{R}^+ / \lambda_n \, \textrm{is eigenvalue}\rbrace,\nonumber\\
&&\textrm{then} \, \forall k_n \in Z(f_{\hat{H}}),	\, k_n^2=\lambda_n \in \tilde{\sigma}(\hat{H}).\,  \textrm{The multiplicity of} \, k_n\,  \textrm{is the}\nonumber\\
&& \textrm{degeneracy of} \, \lambda_n.
\end{eqnarray*}
The formal definition of the spectral zeta function associated to $\hat{H}$ is
\begin{equation}
\zeta_{\hat{H}}(s)= \sum_{\tilde{\sigma}(\hat{H})} \lambda_n^{-s} \qquad \textrm{for Re(s)$>$ certain positive real number} .
\end{equation}
Taking into account that the function
\begin{equation}
\frac{d}{d z}\log ( f_{\hat{H}}(z))
\end{equation}
has poles at $ Z(f_{\hat{H}})$ and that the residue coincides with the multiplicity of the corresponding zero, the summation over $\lambda_n$
 is equivalent to the summation over the zeroes of $f_{\hat{H}}(z)$ and therefore can be written as
 \begin{equation}
 \sum_{\tilde{\sigma}(\hat{H})}\{...\} =\sum_{Z(f_{\hat{H}})}\{...\}= \oint_C dz \frac{d}{d z}\log ( f_{\hat{H}}(z))\{...\}
 \end{equation}
 where $C$ is a contour that encloses all the zeroes contained in $Z(f_{\hat{H}})$. Since $\hat{H}$ is an elliptic non-negative selfadjoint, second order differential operator we can ensure $Z(f_{\hat{H}})\subset\mathbb{R}$. Hence we can choose $C$ to be the semicircle in the complex plane $[-i R,iR]\cup\{z\in\mathbb{C}/\,\,\vert z\vert=R,\,\,{\rm and}\,\,{\rm arg}(z)\in[-\pi,\pi]\}$ and then deform the contour taking the limit $R\to\infty$. After the limit is done, and with the properties assumed for $f_{\hat{H}}(z)$ we obtain an expression for the spectral zeta function that admits analytical continuation to the whole complex plane:
\begin{equation}
\zeta_{\hat{H}}(s)=\frac{\sin(\pi s)}{\pi} \int_0^\infty dk k^{-2s} \partial_k \log [f_{\hat{H}}(ik)].
\end{equation}
In this representation the information about the poles of  $\zeta_{\hat{H}}(s)$ and the values at $s \in \mathbb{Z} $ is contained in
\begin{equation}\label{eq3}
\frac{\sin(\pi s)}{\pi} \int_1^\infty dk k^{-2s} \partial_k \log [f_{\hat{H}}(ik)].
\end{equation}
Hence it all reduces to study \eqref{eq3} in order to obtain the pole structure (Res) and $\zeta_{\hat{H}}(s\in \mathbb{Z})$. In section 3.2 of Ref. \cite{mc-kk1} it can be seen an example where all the calculations can be performed analytically.

\section{The comb as a piston}
In order to perform the calculation of the quantum vacuum energy per unit cell for the comb, it is of great interest to re-interpret the corresponding quantum system as a one-parameter family of hamiltonians defined over the finite interval, by using general quantum boundary conditions in the formalism described in \cite{aso-mc1,aso-mc2}. Bloch's theorem ensures that knowing the wave functions on a primitive cell is equivalent to the knowledge of the wave function in the whole lattice. Hence, if the origin of the real line is chosen in a way that it is coincident with one of the lattice potential centres, then it is enough to study the quantum mechanical system characterised by the quantum hamiltonian
\begin{equation}\label{eq8}
H=-\der{}{x}{2} + w_0 \delta(x) +2w_1 \delta^{\prime}(x),
\end{equation}
defined over the closed interval $[-a/2,a/2]$, being $a$ the lattice spacing. Since the hamiltonian in \eqref{eq8} is not essentially selfadjoint when is defined over the square integrable functions over the closed interval $[-a/2,a/2]$ we need to impose boundary conditions at $x=\pm a/2$ over the boundary values $\{\psi(\pm a/2),\psi'(\pm a/2)\}$. If in addition such boundary condition ensures that the domain of the corresponding selfadjoint extension is a set of wave functions that satisfy Bloch's semi-periodicity condition\footnote{It is of note that in the interval  $[-a/2,a/2]$, the subinterval  $[-a/2,0)$ belongs to one primitive cell, meanwhile the subinterval  $(0,a/2]$ belongs to a different primitive cell.}, then we can understand the comb as a 1-parameter family of selfadjoint extensions where the parameter is to be interpreted as the quasi-momentum. Below we construct the family of selfadjoint extensions that model the comb.

To start with, let us study the $\delta$-$\delta^\prime$ potential sitting at $x=0$ and confined in the interval [-a/2, a/2]. The hamiltonian of the system is given by \eqref{eq8}
%\begin{equation}\label{eq8}
%H=-\der{}{x}{2} + w_0 \delta(x) +2w_1 \delta^{\prime}(x)
%\end{equation}
and its domain (the space of quantum states) in general would be characterised by the general boundary condition
\begin{eqnarray}\label{eq9}
 \begin{pmatrix} \psi(-a/2)+i\psi^\prime(-a/2)  \\ \psi(a/2)-i\psi^\prime(a/2)  \end{pmatrix}=U\begin{pmatrix}  \psi(-a/2)-i\psi^\prime(-a/2)  \\  \psi(a/2)+i\psi^\prime(a/2)  \end{pmatrix},
\end{eqnarray}
where $U\in SU(2)$. In general any $U\in SU(2)$ makes (\ref{eq8}) selfadjoint in the interval $[-a/2, a/2]$. Nevertheless we are focused on mimicking with (\ref{eq9})  Bloch's semi-periodicity condition:
\begin{eqnarray}\label{eq10}
\psi(a/2)&=&e^{iqa} \psi(-a/2)\nonumber\\
\psi^\prime(a/2)&=&e^{iqa} \psi^\prime(-a/2).
\end{eqnarray}
It is straightforward to see that the $U$ that gives rise to (\ref{eq10}) is given by
\begin{eqnarray}\label{eq11}
U_B= \begin{pmatrix} 0 & e^{i\theta} \\ e^{-i\theta} & 0  \end{pmatrix}.
\end{eqnarray}
Plugging (\ref{eq11}) in (\ref{eq9}) one gets
\begin{eqnarray}
\psi(a/2)+i\psi^\prime(a/2) &=& e^{-i\theta}[ \psi(-a/2)+i\psi^\prime(-a/2)]\nonumber\\
 \psi(a/2)-i\psi^\prime(a/2) &=& e^{-i\theta} [\psi(-a/2)-i\psi^\prime(-a/2) ].
\end{eqnarray}
Adding and subtracting both expressions we obtain
\begin{eqnarray}
\psi(a/2)&=&e^{-i\theta} \psi(-a/2)\nonumber\\
\psi^\prime(a/2)&=&e^{-i\theta} \psi^\prime(-a/2),
\end{eqnarray}
and making $\theta=-qa$, we obtain the expressions (\ref{eq10}). Hence, the selfadjoint extension that gives Bloch's condition is given by
\begin{eqnarray}\label{eq15}
U_B=\begin{pmatrix} 0&e^{-iqa}\\ e^{iqa}&0 \end{pmatrix}.
\end{eqnarray}
In addition let us remember that the matching conditions that define the potential ${V=w_0\delta(x)+2w_1\delta^{\prime}(x)}$ are given by (see ref. \cite{GNN})
\begin{eqnarray}\label{eq16}
\begin{pmatrix} \psi(0^+)\\\psi^\prime(0^+) \end{pmatrix}=\begin{pmatrix} \alpha &0 \\ \beta &1/ \alpha \end{pmatrix} \begin{pmatrix} \psi(0^-)\\\psi^\prime(0^-) \end{pmatrix} \qquad \alpha=\frac{1+w_1}{1-w_1}, \qquad \beta=\frac{w_0}{1-w_1^2}.
\end{eqnarray}
When we solve the equation:
\begin{equation}
-\der{}{x}{2} \psi(x)=k^2 \psi(x),
\end{equation}
with the matching conditions (\ref{eq16}) and the boundary condition (\ref{eq9}) with $U=U_B$  given in (\ref{eq15}) we can rearrange everything to write down the secular equation and the general solution in terms of the scattering data for the $\delta$-$\delta'$ potential over the real line as was done in Ref. \cite{course-comb}. This approach enables to interpret the $\delta$-$\delta^\prime$ comb as a one-parameter family of quantum pistons by reinterpreting the primitive cell of the comb in the following way:
\begin{enumerate}
\item The middle piston membrane is represented by the $\delta$-$\delta^\prime$ potential placed at $x=0$. To ensure that the lattice quantum fields satisfy the matching conditions \eqref{eq16} we can assume the ansatz for the one-particle states wave functions in $[-a/2,a/2]$ is given by a linear combination of the two linear independent scattering states determined by the scattering amplitudes of the $\delta$-$\delta^\prime$ potential (see Refs. \cite{GNN,MM,gadella-jpa16})
\begin{equation}\label{scat-amp}
t=\frac{-2k(w_1^2-1)}{2k(w_1^2+1)+iw_0}, \quad \!\!\! r_R=\frac{-4kw_1-iw_0}{2k(w_1^2+1)+iw_0}, \quad \!\!\! r_L=\frac{4kw_1-iw_0}{2k(w_1^2+1)+iw_0}.
\end{equation}
From this amplitudes the determinant of the scattering matrix reads
\begin{equation}
\Rightarrow\textrm{det}\, S_{\delta \delta^{\prime}}=t^2-r_Rr_L=\frac{2k(w_1^2+1)-iw_0}{2k(w_1^2+1)+iw_0}.
\end{equation}
\item The endpoints of the primitive cell correspond to the external walls of the piston placed at $x=\pm a/2$, and the quantum field satisfy the one-parameter family of quantum boundary conditions depending on the parameter $\theta=q a$, which is the quasi-momentum, given by the unitary matrix $U_B$ in \eqref{eq11}.
\end{enumerate}

The spectral function for $U=U_B$ written in terms of the scattering data ($t, r_R, r_L$) and the quasi-momentum $q$ is (see formula (34) in ref. \cite{course-comb})
\begin{equation}\label{eq18}
h(k)=4k\, [2t\cos(qa)-e^{-ika}-e^{ika}(t^2-r_R r_L)].
\end{equation}
The band structure of this comb is given by those $k_j$ such that $h(k_j)=0$. In general the solutions $\{k_0, k_1,  ...,k_n,...\}$ are functions of $q\in[-\pi/a,\pi/a]$, so $k_j(q)^2$ is an energy band when we let $q$ take its continuum values  in $[-\pi/a,\pi/a]$.
In order to use zeta function regularisation we need to remove in \eqref{eq18} the $4k$ global factor to get the ``good" spectral function according to section 2 (see Refs. \cite{spectral,mc-kk1}). Hence the spectral function to be used in our zeta function regularisation approach is given by
\begin{equation}\label{eq19}
f_q(k)=2t \left[\cos(qa)-\frac{1}{2t}(e^{-ika}+e^{ika}(t^2-r_R r_L))\right].
\end{equation}
In \eqref{eq18} and \eqref{eq19}, $t,r_R,r_L$ are the scattering data for the compact supported potential from which the comb is built up on the real line. In addition it is trivial to see that
\begin{equation}
f_q(k)=0 \quad \rightarrow \quad \cos(qa)=\frac{1}{2t}[e^{-ika}+e^{ika}(t^2-r_R r_L)],
\end{equation}
which is the usual form for the band equation written in standard text books such as \cite{ashcroft}, and generalised in \cite{course-comb}.
 Note that because $t^2-r_R r_L$ is the determinant of the unitary scattering matrix, then $t(0)^2-r_R(0)r_L(0)\neq 0$. Hence, in general we can work under the assumption that
\begin{equation}
\lim_{k\to 0} f_q(k)\neq 0,\infty .
\end{equation}

\paragraph{REMARK.} It is of note that all the formulas presented in this section, specially \eqref{eq19} is valid for any comb built from repetition of potentials with compact support smaller than the lattice spacing. All that is needed are the scattering amplitudes for a single potential of compact support over the real line, to obtain the corresponding spectral function that characterises the band structure of the corresponding comb.

\section{Spectral zeta function for the crystal}
Following the interpretation of the comb as a 1-parameter family of selfadjoint extensions given in the previous section we rethink the band spectrum in the following way
\begin{enumerate}
\item For a fixed value of $q\in [-\pi/a, \pi/a]$, $f_q(k)=0$ with $f_q(k)$ given by (\ref{eq19}), gives a discrete set of values of $k$ in one-to-one correspondence with $\mathbb{N}$.
\item  If we let $q$ take values from $-\pi/a$ to $\pi/a$ and put together all the discrete spectra from the previous item, then we will obtain all the allowed energy bands.
\end{enumerate}
Hence in order to perform the calculation of the quantum vacuum energy for a massless scalar field we can write down the spectral zeta function that corresponds to the Schr\"odinger Hamiltonian of the comb
\begin{equation}
\sum_{bands} \int_{\sqrt{\epsilon_{min}^{(n)}}}^{\sqrt{\epsilon_{max}^{(n)}}} dk k^{-2s} = \int_{-\pi/a}^{\pi/a} \frac{dq\,  a}{2\pi}\frac{\sin(\pi s)}{\pi} \int_0^{\infty} dk k^{-2s} \partial_k \log f_q(ik).
\end{equation}
In this way we can write in general the spectral zeta function for the comb as
\begin{equation}
\zeta_{C}(s)= \frac{a}{2\pi}\int_{-\pi/a}^{\pi/a} d q\frac{\sin(\pi s)}{\pi} \int_0^{\infty} dk k^{-2s} \partial_k \log f_q(ik).
\end{equation}
Since the integration in $q$ runs over a finite interval, and $q$ enters as a parameter of the selfadjoint extension associated to the unitary operator $U_B$ in \eqref{eq15}, all the infinite contributions of the quantum vacuum energy are enclosed in the zeta function for a $\delta$-$\delta^\prime$ potential placed at $x=0$ confined between two plates placed at $x=\pm a/2$, i. e.
\begin{equation}
\zeta_q(s)=\frac{\sin(\pi s)}{\pi} \int_0^{\infty} dk k^{-2s} \partial_k \log f_q(ik).
\end{equation}
As a result of the formulas for the spectral zeta function it is easy to conclude that the finite quantum vacuum energy for the comb, $E_{\rm comb}^{fin}$, can be obtained from the finite quantum vacuum energy $E_0^{fin}(q)$ for the quantum scalar field confined between two plates placed at $x=\pm a/2$ represented by the boundary condition associated to \eqref{eq15}, and under the influence of a $\delta$-$\delta^\prime$ potential placed at $x=0$:
\begin{equation}
E_{\rm comb}^{fin}=\frac{ a}{2\pi}\int_{-\pi/a}^{\pi/a}d qE_0^{fin}(q).
\end{equation}
Hence our problem reduces to compute $E_0^{fin}(q)$.

\section{The finite quantum vacuum energy at zero temperature for generalised Dirac combs.}

\subsection{General formulas}
From this point we will use formula 2.26 in Ref. \cite{aso-mc2} to obtain $E_0^{fin}(q)$. In Ref. \cite{aso-mc2} there was no point potential between plates, so the final result arising there did not depend on the reference length $L_0$ used to subtract the infinite parts. In our case the existence of a potential with compact support between plates forces to take the limit $L_0\to\infty$. Physically this limit means that what we subtract is the quantum vacuum energy of the potential with compact support on the whole real line. With these assumptions and changing the length $L$ in Ref. \cite{aso-mc2} by our lattice spacing $a$ we can write
\begin{equation}
E_0^{fin}(q)=\lim_{a_0\to\infty}\frac{- a_0}{2\pi(a-a_0)}\int_0^{\infty} dk \, k \left[ a-a_0-\deruno{}{k} \log \left(\frac{f_q^a(ik)}{f_q^{a_0}(ik)}\right) \right].
\end{equation}
In taking this limit, we must keep $q a=q a_0=-\theta$ as a free parameter coming from the selfadjoint extension, and just after having done the limit and obtained a finite result make the replacement $\theta=-q a$. Hence to avoid confusion we can write
\begin{equation}\label{e-fin-theta}
E_0^{fin}(\theta)=\lim_{a_0\to\infty}\frac{- a_0}{2\pi(a-a_0)}\int_0^{\infty} dk \, k \left[ a-a_0-\deruno{}{k} \log \left(\frac{f_\theta^a(ik)}{f_\theta^{a_0}(ik)}\right) \right],
\end{equation}
with
\begin{equation}\label{eq19-theta}
f^a_\theta(k)=2t\left[\cos(\theta)-\frac{1}{2t}(e^{-ika}+e^{ika}(t^2-r_R r_L))\right],
\end{equation}
and finally
\begin{equation}\label{efin-comb}
E_{\rm comb}^{fin}=\int_{-\pi}^{\pi}\frac{d\theta}{2\pi}E_0^{fin}(\theta),
\end{equation}
being $\theta$ the parameter of the selfadjoint extension defined by $U_B$ that is to be interpreted after obtaining a finite answer as $\theta=-q a$.
\subsection{Some comments on $E_{\rm comb}^{fin}$ and $E_0^{fin}(\theta)$.}
With the formulas written above for the finite quantum vacuum energy of the comb ($E_{\rm comb}^{fin}$) and the finite quantum vacuum interaction energy between two plates modelled by the boundary condition associated to $U_B$ with a compact supported potential centred in the middle point of both plates ($E_0^{fin}(\theta)$), we are assuming that the zero point energy corresponds to the situation in which we have a free scalar quantum field over the real line. Under this assumption when the potential with compact support between plates is made identically zero ($t=1,r_R=r_L=0$), the quantity
\begin{equation}
{\cal E}_0(\theta)\equiv\left. E_0^{fin}(\theta)\right\vert_{t=1,r_R=r_L=0}\neq 0,\infty,
\end{equation}
is nothing but the scalar quantum vacuum interaction energy between two plates mimicked by quasi-periodic boundary conditions. This was analytically obtained in Refs. \cite{aso-mc2,ijtp-asomc11} for the 1D, 2D and 3D cases. The fact that ${\cal E}_0(\theta)\neq 0,\infty$ means that one would expect
\begin{equation}\label{efin-comb2}
E_{\rm comb}^{fin}(t=1,r_R=r_L=0)=\int_{-\pi}^{\pi}\frac{d\theta}{2\pi}{\cal E}_0(\theta)\neq 0,\infty,
\end{equation}
which makes sense, since turning off the potential with compact support does not leave us with a quantum scalar field over the real line, because the Bloch periodicity condition remains. Nevertheless if we take into account that any plane wave on the real line satisfies Bloch periodicity, the energy $E_{\rm comb}^{fin}(t=1,r_R=r_L=0)$ should be that of the free scalar field on the real line, i. e. zero. Knowing from Refs. \cite{ijtp-asomc11,phd-jmmc} that
\begin{equation*}
{\cal E}_0(\theta)=\frac{1}{2a}\left( \vert\theta\vert-\frac{\theta^2}{2\pi}-\frac{\pi}{3}\right),
\end{equation*}
it is straightforward to see that
\begin{equation}
E_{\rm comb}^{fin}(t=1,r_R=r_L=0)=\int_{-\pi}^{\pi}\frac{d\theta}{2\pi}{\cal E}_0(\theta)=0.
\end{equation}
As a result, we ensure that our general formula \eqref{efin-comb} gives total quantum vacuum energy for the comb identically zero when the potentials with compact support that form the comb are zero, as it should be.
\subsection{$E_{\rm comb}^{fin}$ for the $\delta$-$\delta^\prime$ comb.}

\begin{figure}[H]
\centering
\includegraphics[scale=0.65]{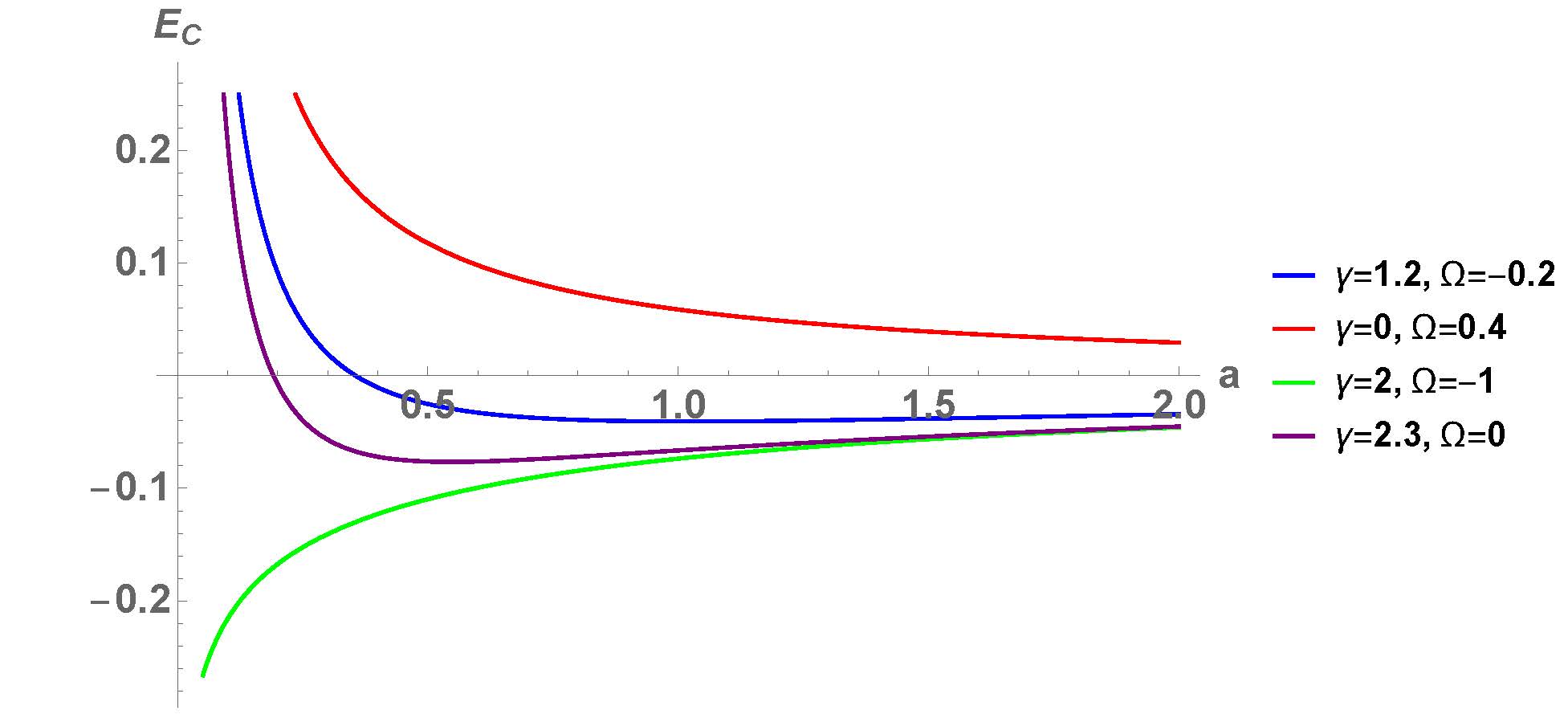}
\caption{\small Quantum vacuum energy as a function of the distance $a$ for different values of the $\delta \delta^{\prime}$ couplings. }
\label{fig:en2}
\end{figure}

Plugging the scattering amplitudes given in \eqref{eq19} and after some algebraic manipulations we obtain
\begin{equation}\label{fddp}
f_\theta(k)=-\frac{4k(1+w_1^2)}{\Delta_{\delta\delta^\prime}}\left[\Omega \cos(\theta)+\cos(ka)+\frac{\gamma}{2k} \sin(ka)\right],
\end{equation}
being $\Delta_{\delta\delta^\prime}=[2k(w_1^2+1)+iw_0]^2$, $\Omega\equiv(w_1^2-1)/(w_1^2+1)$ and $\gamma\equiv w_0/(1+w_1^2)$. In order to have a well behaved spectral function ($f_\theta(k\to 0)\neq0$) we have to remove the global $-4 k(1+w_1^2)$ factor. In addition, the global factor $1/\Delta_{\delta\delta^\prime}$ does not change the zeroes of the spectral function so it can also be dropped. Hence we obtain the following expression for the spectral function of the $\delta$-$\delta^\prime$ comb:
\begin{equation}\label{eq32}
g_\theta(k)=\Omega\cos(\theta)+\cos(ka)+\frac{\gamma}{2k} \sin(ka).
\end{equation}
The quantum vacuum energy is obtained from equation \eqref{efin-comb} after taking the limit ${a_0\to\infty}$:
\begin{equation}\label{38}
E_{\delta\delta'{\rm comb}}^{fin}=\int_{-\pi}^{\pi}\frac{d\theta}{4\pi^2}\int_0^{\infty} dk \, F_{\delta\delta'}(k,\theta),
\end{equation}
where
\begin{equation}\label{eq34}
 F_{\delta\delta'}(k,\theta)=\frac{A(k)}{B(k)+C(k) \cos \theta } + ak -\frac{\gamma}{\gamma + 2k},
\end{equation}
and $A(k)$, $B(k)$ and $C(k)$  are defined as
\begin{eqnarray}
&&\hspace{-0.5cm}A(k)=-ak\gamma \cosh (ka) + (-2ak^2 + \gamma ) \sinh(ka)  \label{a-b}\\
&&\hspace{-0.5cm} B(k)=2k \cosh(ka) + \gamma \sinh (ka),\quad C(k)= 2k\Omega.  \label{ec-c}
\end{eqnarray}
Since now everything is finite in \eqref{38} we can exchange the order of integration to do first the integration in $\theta$
\begin{equation}\label{int-theta}
I_{\delta\delta'}(k)=\int_{-\pi}^{\pi}\frac{d\theta}{4\pi^2} \, F_{\delta\delta'}(k,\theta).
\end{equation}
The integral \eqref{int-theta} can be obtained from Ref. \cite{gradshteyn} (page 402 formula 3.645)
\begin{eqnarray*}
&&\int_0^\pi \frac{\cos^n(x)}{(b+a\cos x)^{n+1}}=\frac{\pi}{2^n (b+a)^n \sqrt{b^2-a^2}} \\
&&\times \sum_{k=0}^n (-1)^k \frac{(2n-2k-1)!!(2k-1)!!}{(n-k)!k!} \left(\frac{a+b}{b-a}\right)^k,
\end{eqnarray*}
for $b^2>a^2$. In order to use this integral to obtain $I(k)$ in \eqref{int-theta} we need to ensure that $B^2(k,a)>C^2(k,a)$. Taking into account the definition of $B(k), C(k)$ in \eqref{ec-c},  this condition is always fulfilled because  $-1<\Omega<1$ and
\begin{equation}
 \cosh (ka) + \frac{\gamma}{2k} \sinh (ka) > 1, \quad \forall k,a, \gamma >0.
\end{equation}
Hence the integration in $\theta$ is given by
\begin{equation}\label{i-of-k}
I_{\delta\delta'}(k)=\frac{1}{2\pi}\left[ \frac{A(k)}{\sqrt{B^2(k)-C^2(k)}}+ ak -\frac{\gamma}{\gamma+2k}.\right]
\end{equation}
With this result the quantum vacuum energy for the comb is finally reduced to a single integration in $k$ :
\begin{equation}\label{e-final}
E_{\delta\delta'{\rm comb}}^{fin}=\int_0^\infty dk I_{\delta\delta'}(k).
\end{equation}
This integral can be calculated numerically with \textit{Mathematica}. The results are shown below. As can be seen in Fig. \ref{fig:en1} the quantum vacuum energy produced by a quantum scalar field can be positive (repulsive force), negative (attractive force) or zero. Taking into account that the potentials sitting in each lattice node mimic atoms that have lost their most external electron, classically the force between them is repulsive (they all have positive charge). The fact that the quantum vacuum energy of the scalar field can be negative and hence reduce the repulsive classical force means that when the quantum vacuum energy is negative the lattice spacing tends to be smaller. On the other hand when the quantum vacuum force is positive the classical repulsion is enhanced promoting that the lattice spacing in the crystal becomes bigger. Figure \ref{fig:en2} shows the behaviour of the quantum vacuum energy \eqref{e-final} as a function of the lattice spacing $a$. In all the cases shown the quantum vacuum energy becomes zero as $a\to\infty$ and tends to $\pm\infty$ as $a\to 0$. In addition it is very easy to check that in the limit $\gamma\to\infty$, i.e. $w_0\to\infty$,
\begin{equation}
\lim_{\gamma\to\infty}I_{\delta\delta'}(k)=-\frac{ka \, \, e^{-ka}\, \text{csch}(ka)}{2 \pi  },
\end{equation}
one recovers the very well known result for the quantum vacuum energy between two Dirichlet plates in $1+1$: $E_0=-\pi/(24 a)$. The limit $w_0\to\infty$ gives the minimum quantum vacuum energy that the $\delta$-$\delta'$ can have. On the other hand from  Fig. \ref{fig:en1} it is easy to see that the maximum energy is positive, and it occurs for $\Omega=\gamma=0$, i.e. $w_1=\pm1$ and $w_0=0$. In this case
\begin{equation}
\lim_{\Omega,\gamma\to 0}I_{\delta\delta'}(k)=-\frac{k a (\tanh (k a)-1)}{2 \pi  }\Rightarrow E_{\delta\delta'{\rm comb}}^{fin}(\Omega=\gamma=0)= \frac{\pi }{48 a},
\end{equation}
and it corresponds to mixed boundary conditions \cite{AmbWol-AOP83,ElizRom-PRD89}, where Dirichlet boundary conditions are imposed on one side and Neumann ones on the other.

\begin{figure}[h]
\centering
\includegraphics[scale=0.65]{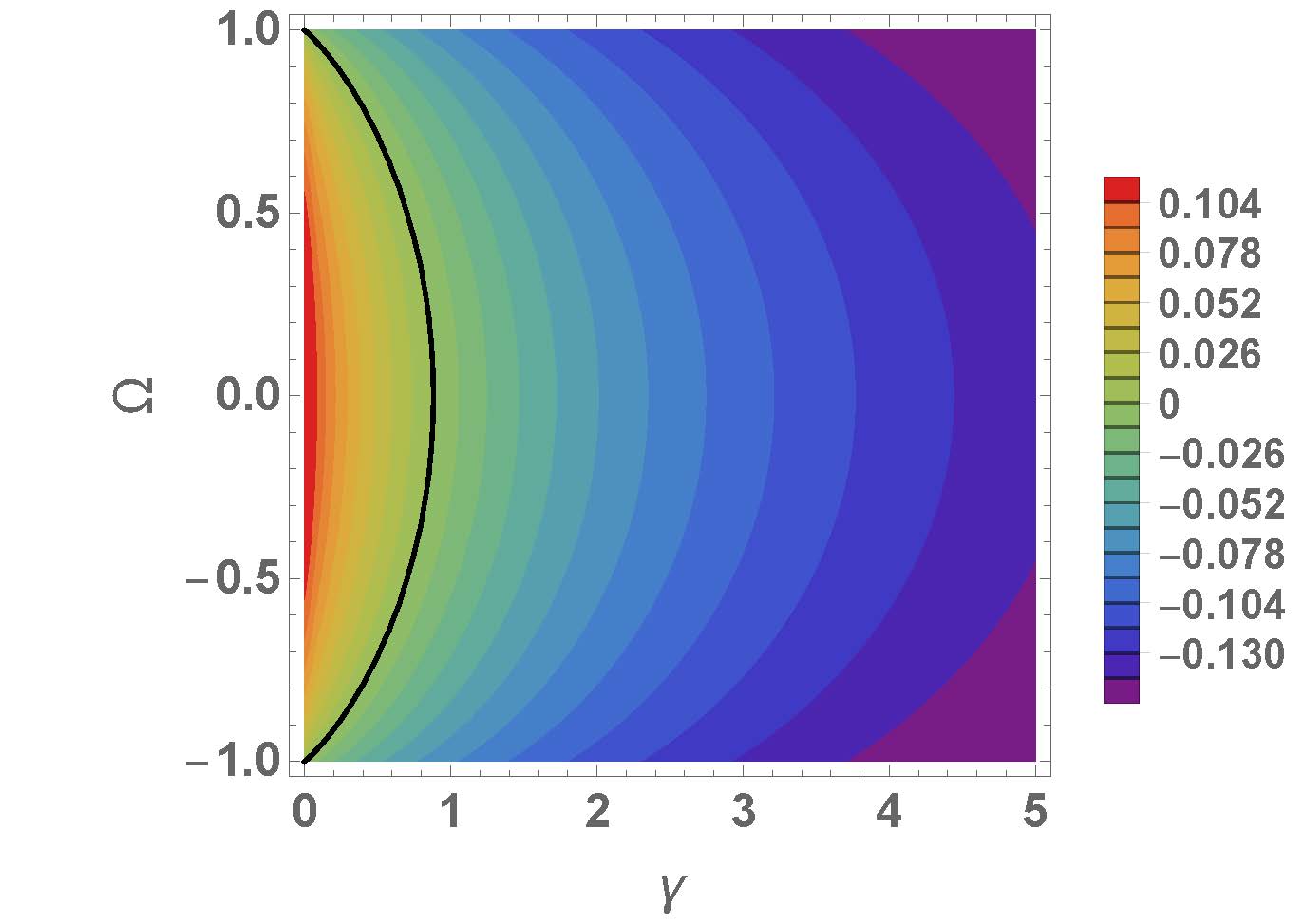}
\caption{\small Quantum vacuum energy for $a=0.5$ in the coupling space $\gamma$-$\Omega$.}
\label{fig:en1}
\end{figure}

It is interesting to remark that, as it happens for the quantum vacuum interaction energy between two Dirac-
$\delta$ plates in a $1+1$ dimensional scalar quantum field theory, the limit
\begin{equation}
\lim_{w_0\to 0}E_{\delta\delta'{\rm comb}}^{fin}(\Omega=-1,w_0),
\end{equation}
is not analytical in $w_0$ due to the infrared divergence that appears in the Feymann diagrams (see refs. \cite{toms-jpa45,milton-jpa04}). This can be seen in eq. \Ref{i-of-k} if we take into account that the non analyticity is enclosed in the third term of the r.h.s.

\section{Conclusions and further comments}
We calculated the quantum vacuum energy of a comb formed by linear combinations of $\delta$- and $\delta'$-functions given in eq. \Ref{1}. The method presented in this paper is based on the spectral zeta function. We showed that the $\delta$-$\delta'$ comb with lattice spacing $a$ is equivalent to a single $\delta$-$\delta'$ potential in the interval $[-a/2,a/2]$ at $x=0$ together with a 1-parameter family of quasi-periodic boundary conditions at $x=\pm a/2$ given by eq. \Ref{eq15}. The band structure, eq. \Ref{eq18},  arises when one takes into account that the spectrum of the comb is the set obtained by the union of all the discrete spectra of all the selfadjoint extensions obtained from the 1-parameter family of boundary conditions \Ref{eq15}. The method can be easily generalised to any comb formed by the repetition of potentials with compact support, as long as the compact support is smaller than the lattice spacing. The ultraviolet divergences of these combs are the same as those of the quantum vacuum energy for one potential with compact support over the real line, which does not have a band structure but a continuum spectrum. Therefore the ultraviolet divergences for the kind of combs studied in this paper do not depend on the lattice spacing. Subtracting these contributions we get a finite quantum vacuum energy that represents the part of the vacuum expectation value of the Hamiltonian of the quantum field theory, which depends on the lattice spacing. As expected, the generalised vacuum energy vanishes in the limit of infinite lattice spacing. This procedure has already been applied in \cite{bord92-25-4483} for two $\delta$-functions. The interpretation of this vacuum energy is a contribution (one-loop quantum correction) to the elastic lattice forces produced by the quantum scalar field of the phonons.

\begin{figure}[H]
\centering
\includegraphics[scale=0.50]{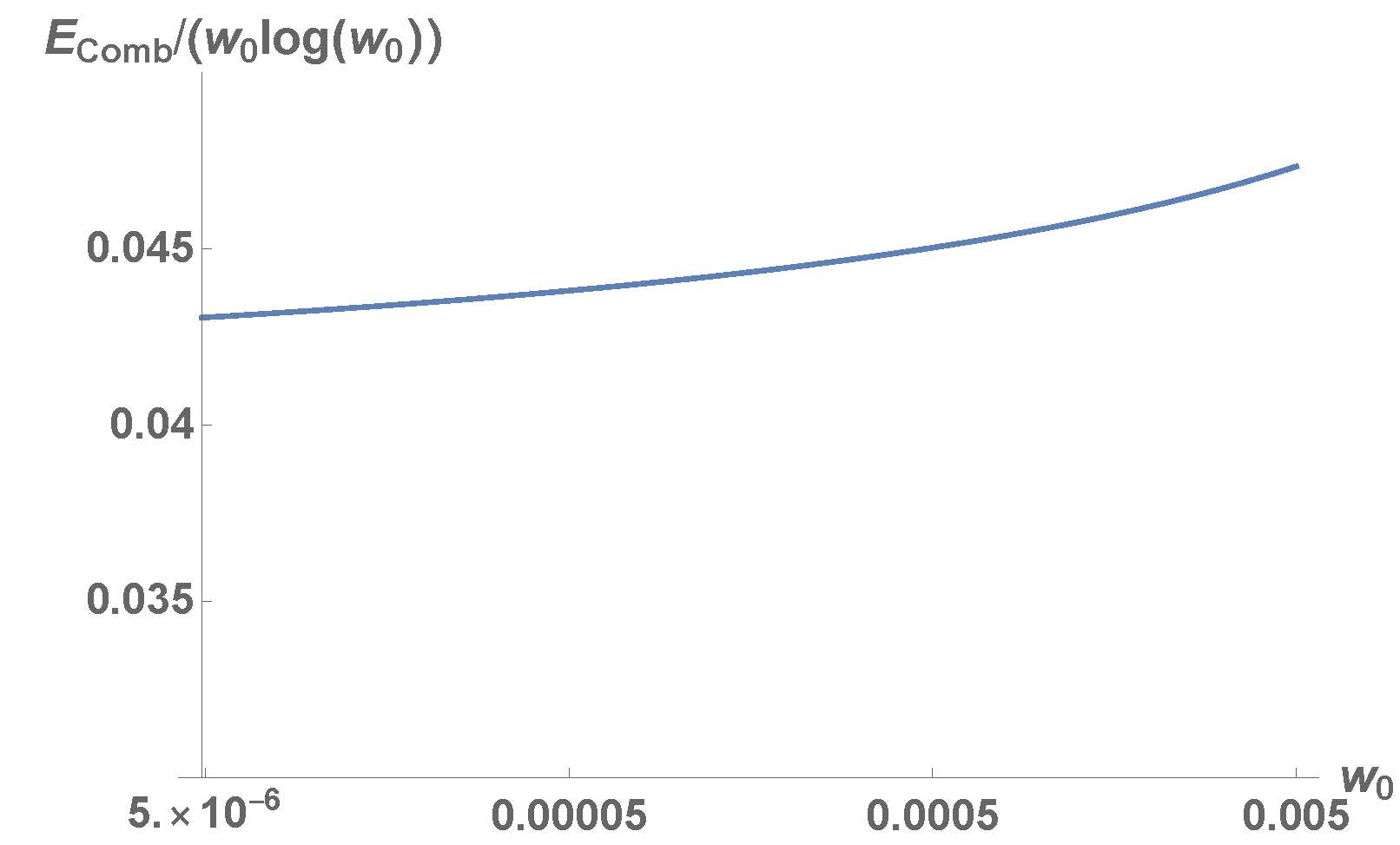}
\caption{\small Plot of $E_{\delta\delta'{\rm comb}}^{fin}(\Omega=-1,w_0\to 0)/( w_0\log{w_0})$ for $a=1$. The $w_0$ axis is in logarithmic scale }
\label{fig:en3}
\end{figure}

The calculations are to a large extent explicit. The result, eq. \Ref{e-final}, has a fast converging single integration over $k$ with the integrand \Ref{i-of-k}, given in terms of elementary functions: exponential and hyperbolic functions. This integration is over imaginary frequency after performing a Wick rotation \cite{bord92-25-4483}. In addition, the result presented in eq.  \Ref{e-final} enables us to infer that when $w_1=0$, i.e. $\Omega=-1$, the function $E_{\delta\delta'{\rm comb}}^{fin}(\Omega=-1,\gamma=w_0)$ is not analytical when $w_0\to 0$. Moreover, the plot in Fig. \ref{fig:en3} shows that 
\begin{equation}
E_{\delta\delta'{\rm comb}}^{fin}(\Omega=-1,w_0\to 0)\sim w_0\log{w_0},
\end{equation}
as known from the vacuum energy of a single delta function (see refs. \cite{toms-jpa45,milton-jpa04}). 
\noindent From this we can conclude that $E_{\delta\delta'{\rm comb}}^{fin}(w_1,w_0)$ does not admit a perturbative expansion in powers of $w_0$ around $w_0=0$ when $w_1=0$. Hence the result given in formula \Ref{e-final} is non-perturbative in the sense that there is no power series expansion for $E_{\delta\delta'{\rm comb}}^{fin}(\Omega=-1,w_0)$ when $w_0\to 0$.

With a two-dimensional parameter space (the strength $w_0$ of the $\delta$-potential and the strength $w_1$ of the $\delta'$-potential) the quantum vacuum energy can be positive (repulsive force between nodes of the lattice) and negative (attractive force between nodes of the lattice). The interface between the two regimes mentioned is the line of zero quantum vacuum energy in the $\Omega$-$\gamma$ plane shown in Fig. \ref{fig:en1}.

The techniques developed in this paper have provide a framework to calculate relevant quantities such as the free energy and the entropy at finite temperatures different from zero.

\section*{Acknowledgements}
The authors acknowledge support from the German Research Foundation (DFG) and Universit\"at Leipzig within the program of Open Access Publishing. JMMC and LSS are grateful to the Spanish Government-MINECO (MTM2014-57129-C2-1-P) and the Junta de Castilla y Le\'on (BU229P18, VA137G18 and VA057U16) for the financial support. JMMC would like to thank J. Mateos Guilarte, M. Gadella and L. M. Nieto for fruitful discussions. JMMC dedicates this paper to the memory of professor Jose M. Mu$\tilde{\rm n}$oz Porras.

\bibliographystyle{unsrt}

\bibliography{qvacuum_energycomb-bib}{}

\begin{thebibliography}{10}

\bibitem{kronig-prsa31}
R.~de~L.~Kronig and W.~G. Penney.
\newblock Quantum mechanics of electrons in crystal lattices.
\newblock {\em Proceedings of the Royal Society of London A: Mathematical,
  Physical and Engineering Sciences}, 130(814):499--513, 1931.

\bibitem{GNN}
M.~Gadella, J.~Negro, and L.M. Nieto.
\newblock Bound states and scattering coefficients of the
  $-a\delta(x)+b\delta^\prime(x)$ potential.
\newblock {\em Physics Letters A}, 373(15):1310 -- 1313, 2009.

\bibitem{MM}
J.~M. Mu$\tilde{\rm n}$oz-Casta$\tilde{\rm n}$eda and J.~Mateos Guilarte.
\newblock {$\delta$-$\delta^\prime$ generalized Robin boundary conditions and
  quantum vacuum fluctuations}.
\newblock {\em Phys. Rev.}, D91(2):025028, 2015.

\bibitem{gadella-jpa16}
M~Gadella, J~Mateos-Guilarte, J~M Mu$\tilde{\rm n}$oz-Casta$\tilde{\rm n}$eda,
  and L~M Nieto.
\newblock Two-point one-dimensional $\delta$-$\delta^\prime $ interactions:
  non-abelian addition law and decoupling limit.
\newblock {\em Journal of Physics A: Mathematical and Theoretical},
  49(1):015204, 2016.

\bibitem{FAS}
Sergio Albeverio, Silvestro Fassari, and Fabio Rinaldi.
\newblock A remarkable spectral feature of the schr\"odinger hamiltonian of the
  harmonic oscillator perturbed by an attractive $\delta^\prime$-interaction
  centred at the origin: double degeneracy and level crossing.
\newblock {\em Journal of Physics A: Mathematical and Theoretical},
  46(38):385305, 2013.

\bibitem{gad-ijtp11}
M.~Gadella, M.~L. Glasser, and L.~M. Nieto.
\newblock One dimensional models with a singular potential of the type
  $\alpha\delta(x)+\beta\delta^\prime(x)$.
\newblock {\em International Journal of Theoretical Physics}, 50(7):2144--2152,
  2011.

\bibitem{gad-ijtp11-2}
M.~Gadella, M.~L. Glasser, and L.~M. Nieto.
\newblock The infinite square well with a singular perturbation.
\newblock {\em International Journal of Theoretical Physics}, 50(7):2191--2200,
  2011.

\bibitem{course-comb}
L~M Nieto, M~Gadella, J~Mateos Guilarte, J~M Mu$\tilde{\rm
  n}$oz-Casta$\tilde{\rm n}$eda, and C~Romaniega.
\newblock Towards modelling qft in real metamaterials: Singular potentials and
  self-adjoint extensions.
\newblock {\em Journal of Physics: Conference Series}, 839(1):012007, 2017.

\bibitem{JMG}
Jose~M. Mu$\tilde{\rm n}$oz-Casta$\tilde{\rm n}$eda, J.~Mateos Guilarte, and
  A.~Moreno Mosquera.
\newblock {Quantum vacuum energies and Casimir forces between partially
  transparent $\delta$-function plates}.
\newblock {\em Phys. Rev.}, D87:105020, 2013.

\bibitem{spectral}
Klaus Kirsten.
\newblock {\em {Spectral functions in mathematics and physics}}.
\newblock Chapman \& Hall/CRC, Boca Raton, FL, 2001, 2001.

\bibitem{vassilevich-physrept03}
D.V. Vassilevich.
\newblock Heat kernel expansion: user's manual.
\newblock {\em Physics Reports}, 388(5):279 -- 360, 2003.

\bibitem{emil-book}
E.~Elizalde.
\newblock {\em Ten Physical Applications of Spectral Zeta Functions}.
\newblock Lecture Notes in Physics. Springer Berlin Heidelberg, 2012.

\bibitem{aso-mc2}
M.~Asorey and J.M. Mu$\tilde{\rm n}$oz-Casta$\tilde{\rm n}$eda.
\newblock {Attractive and Repulsive Casimir Vacuum Energy with General Boundary
  Conditions}.
\newblock {\em Nucl. Phys.}, B874:852--876, 2013.

\bibitem{mc-kk1}
J.M. Mu$\tilde{\rm n}$oz-Casta$\tilde{\rm n}$eda, Klaus Kirsten, and M.~Bordag.
\newblock {QFT over the finite line. Heat kernel coefficients, spectral zeta
  functions and selfadjoint extensions}.
\newblock {\em Lett. Math. Phys.}, 105(4):523--549, 2015.

\bibitem{aso-mc1}
M.~Asorey, D.~Garcia-Alvarez, and J.M. Mu$\tilde{\rm n}$oz-Casta$\tilde{\rm
  n}$eda.
\newblock {Casimir Effect and Global Theory of Boundary Conditions}.
\newblock {\em J. Phys.}, A39:6127--6136, 2006.

\bibitem{ashcroft}
N.W. Ashcroft and N.D. Mermin.
\newblock {\em Solid State Physics}.
\newblock HRW international editions. Holt, Rinehart and Winston, 1976.

\bibitem{ijtp-asomc11}
M.~Asorey and J.~M. Mu$\tilde{\rm n}$oz-Casta$\tilde{\rm n}$eda.
\newblock Vacuum boundary effects.
\newblock {\em International Journal of Theoretical Physics}, 50(7):2211--2221,
  Jul 2011.

\bibitem{phd-jmmc}
J.~M. Mu$\tilde{\rm n}$oz-Casta$\tilde{\rm n}$eda.
\newblock {\em {Boundary Effects in Quantum Field Theory (in Spanish)}}.
\newblock PhD dissertation, Universidad de Zaragoza, 2009.

\bibitem{gradshteyn}
I.S. Gradshteyn, A.~Jeffrey, and I.M. Ryzhik.
\newblock {\em Table of Integrals, Series, and Products}.
\newblock Academic Press, 1996.

\bibitem{AmbWol-AOP83}
Jan~Ambj\o rn and Stephen Wolfram.
\newblock Properties of the vacuum. i. mechanical and thermodynamic.
\newblock {\em Annals of Physics}, 147(1):1 -- 32, 1983.

\bibitem{ElizRom-PRD89}
E.~Elizalde and A.~Romeo.
\newblock Rigorous extension of the proof of zeta-function regularization.
\newblock {\em Phys. Rev. D}, 40:436--443, Jul 1989.

\bibitem{toms-jpa45}
David~J Toms.
\newblock Renormalization and vacuum energy for an interacting scalar field in
  a $\delta$-function potential.
\newblock {\em Journal of Physics A: Mathematical and Theoretical},
  45(37):374026, 2012.

\bibitem{milton-jpa04}
Kimball~A Milton.
\newblock Casimir energies and pressures for $\delta$-function potentials.
\newblock {\em Journal of Physics A: Mathematical and General}, 37(24):6391,
  2004.

\bibitem{bord92-25-4483}
M.~Bordag, D.~Hennig, and D.~Robaschik.
\newblock {Vacuum energy in quantum field theory with external potentials
  concentrated on planes}.
\newblock {\em J. Phys. A}, A25:4483, 1992.

\end{thebibliography}

\end{document}